\documentclass[aps,prb,showpacs]{revtex4}

\usepackage{graphicx}
\usepackage{amsmath}

\graphicspath{{./IMAGES/}}

\begin{document}
\title{On a solution of the Schr\"{o}dinger equation with a hyperbolic double-well potential}

\author{C. A. Downing}
\email[]{c.a.downing@exeter.ac.uk}
\affiliation{School of Physics, University of Exeter, Stocker Road,
Exeter EX4 4QL, United Kingdom}

\date{Received 30 November 2012; revised manuscript received April 30, 2013}

\begin{abstract}
We report a solution of the one-dimensional Schr\"{o}dinger equation with a hyperbolic double-well confining potential via a transformation to the so-called confluent Heun equation. We discuss the requirements on the parameters of the system in which a reduction to confluent Heun polynomials is possible, representing the wavefunctions of bound states. 
\end{abstract}

\maketitle

\section{\label{intro}Introduction}

A small number of exact solutions to the Schr\"{o}dinger equation were obtained historically in the genesis of quantum mechanics.\cite{Morse, Eckart, Rosen, Teller, Scarf} More recently, other exactly-soluble systems have been found by both traditional means\cite{Zhang, Parfitt} and via the factorization techniques of supersymmetric quantum mechanics (SUSY).\cite{Cooper, Mallow} Conventionally, analytical solutions to the Schr\"{o}dinger equation were found via a reduction to a hypergeometric equation,\cite{Manning} an equation with three regular singular points, however recently a new solution via Heun's differential equation\cite{Ronveaux, Slavyanov, Maier, Fiziev} has been reported.\cite{Hall}

Heun's equation, a Fuchsian equation with four regular singular points, was initially studied by the German mathematician Karl Heun in the late 19th century.\cite{Heun} It has several special or limiting cases of great importance in mathematical physics, namely the Lam\'{e}, Mathieu and spheroidal differential equations.\cite{Abramowitz} However, it is only recently that its use in physics has become increasingly widespread,\cite{Hortacsu} with its solutions being used in works ranging from quantum rings\cite{Loos} to black holes.\cite{Hod}

In this work, we report a class of confining potentials that can be transformed to the confluent Heun equation from the one-dimensional Schr\"{o}dinger equation. One case, of a hyperbolic double-well, allows one to reduce the solution to simple polynomials for special values of the system parameters. The double-well problem has been studied extensively\cite{Davis, Konwent, Nieto} and is of continued interest due to its importance as a toy model, from heterostructure physics\cite{Alferov} to the trapping of Bose-Einstein condensates.\cite{Schumm}

Please consider the following class of potentials, defined by two physical parameters $V_0$ and $d$, shaping the potential depth and width respectively, and with two class parameters, with $q = -2, 0, 2, 4, 6$ defining the 'family' and $p = -2, 0,...~q$
\begin{equation}
\label{intro1}
V(x) = -V_0\frac{\sinh^p(x/d)}{\cosh^q(x/d)}.
\end{equation}
Most notable is the case $(q, p) = (2, 0)$, which describes the well-known P\"{o}schl-Teller potential.\cite{Teller} In fact, all constituent potentials of the families $q = -2, 0, 2$ lead to Schr\"{o}dinger equations of no more than three singular points after a transformation of independent variable to $\xi=1/\cosh^2(x/d)$ or $\zeta = \tanh(x/d)$, and so can be reduced to equations of the hypergeometric type. Termination of the resultant hypergeometric functions leads to analytic expressions for the eigenvalues of the problem. 

However, when $q = 4, 6$ the same transformation variables lead to equations of the confluent Heun type. Now, two termination conditions must be satisfied in order to get a polynomial solution to the Schr\"{o}dinger equation. The $(4, p)$ family cannot be terminated due to the first condition, which imposes a simple relation on the physical parameters arising in the problem. The $(6, p)$ family fulfill the first condition but not the second, more involved condition, which also must hold to abort the Taylor series solution. However, remarkably the special case of $(6, 4)$ is an exception and does admit polynomial solutions for certain values of the physical parameters, which will be the main focus of this work.

The rest of this work is as follows. We give a full derivation of our solution in terms of confluent Heun functions in Sec.~\ref{reduction}, along with details of the power series solution. In Sec.~\ref{bound} we provide the eigenvalues of our hyperbolic double-well problem, along with examples of the wavefunctions found for the first few states. We provide numerical support to our analytic results in Sec.~\ref{disc} and finally, we draw some conclusions in Sec.~\ref{conc}. 

\section{\label{reduction}Reduction to a confluent Heun equation}

\begin{figure}[htbp] 
\centering
\includegraphics[width=0.4\textwidth]{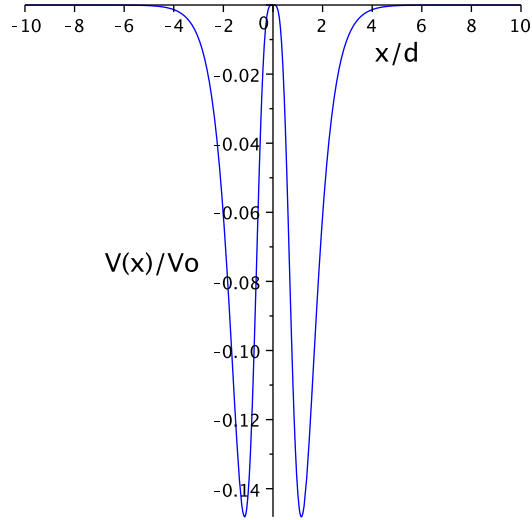}  
\caption{Plot of the hyperbolic double-well potential under consideration, Eq.~\eqref{intro1} with $(q, p)=(6, 4)$.}
\label{fig:pot1}
\end{figure}

The stationary one-dimensional Schr\"{o}dinger equation for a particle of mass $m$ and energy $E$ in a potential $V(x)$ is
\begin{equation}
\label{SE}
\frac{-\hbar^2}{2m} \frac{d^2}{dx^2} \psi(x) + V(x) \psi(x) = E \psi(x).
\end{equation}
We consider the $(6, 4)$ potential from Eq.~\eqref{intro1}, which we display in Fig.~\ref{fig:pot1}, and find upon substitution of $V(x)$
\begin{equation}
\label{SE2}
\frac{d^2}{dz^2} \psi(z) + \left( \varepsilon d^2 + U_0 d^2\frac{\sinh^4(z)}{\cosh^6(z)} \right) \psi(z) = 0,
\end{equation}
where $\varepsilon=2mE/\hbar^2$, $U_0=2mV_0/\hbar^2$ and $z=x/d$. Upon making the change of variable $\xi=1/\cosh^2(z)$, such that the domain $-\infty<x<\infty$ maps to $0<\xi<1$, we find
\begin{equation}
\label{SE3}
\xi^2 (1-\xi)\frac{d^2}{d\xi^2} \psi(\xi) + \xi (1-\frac{3}{2}\xi)\frac{d}{d\xi} \psi(\xi)
 + \frac{1}{4} \left( \varepsilon d^2 + U_0 d^2 \xi (1-\xi)^2 \right) \psi(\xi) = 0.
\end{equation}

Noting that at large $\xi$ Eq.~\eqref{SE3} reduces to
\begin{equation}
\label{SE4}
\frac{d^2}{d\xi^2} \psi(\xi) = \frac{U_0 d^2}{4} \psi(\xi),
\end{equation}
with solutions in the form
\begin{equation}
\label{solution1}
\psi(\xi) = e^{\pm \frac{\alpha}{2}\xi}, \quad \alpha = -d\sqrt{U_0},
\end{equation}
leads us to choose the ansatz solution
\begin{equation}
\label{ansatz1}
\psi(\xi) = e^{\frac{\alpha}{2}\xi}~f(\xi), 
\end{equation}
which yields upon substitution into Eq.~\eqref{SE3}
\begin{equation}
\label{SE5}
\xi^2 (1-\xi)\frac{d^2}{d\xi^2} f(\xi) + \left\{ \alpha \xi^2 (1-\xi) + \xi \left(1-\frac{3}{2}\xi\right)  \right\} \frac{d}{d\xi} f(\xi)
 +  \left\{ \xi^2 (1-\xi) \frac{\alpha^2}{4} + \frac{\alpha}{2} \xi \left( 1- \frac{3}{2} \xi \right) + \frac{\alpha^2}{4}\xi (1-\xi)^2 - \frac{\beta^2}{4} \right\} f(\xi) = 0,
\end{equation}
where $\beta = -i d \sqrt{\varepsilon}$. Undertaking a peeling-off procedure with $f(\xi)=\xi^{\beta/2}~y(\xi)$, we find
\begin{align}
\label{heuneq}
	\frac{d^2}{d\xi^2}y(\xi) +\left( \alpha + \frac{\beta +1}{\xi} + \frac{\gamma +1}{\xi-1} \right) \frac{d}{d\xi}y(\xi) + \left( \frac{ \mu }{\xi} + \frac{\nu}{\xi-1} \right)y(\xi) = 0,
\end{align}
where
\begin{equation*}
\label{coeffs}
	\alpha= -d\sqrt{U_0},\quad\beta = -i d \sqrt{\varepsilon},\quad\gamma=-\frac{1}{2},\quad\mu= \frac{1}{4}\left( \alpha (\alpha + 2) + 2\alpha \beta- \beta (\beta+1) \right),\quad\nu= \frac{1}{4}\left( \alpha + \beta (\beta + 1) \right).
\end{equation*}
This is Heun's confluent differential equation.\cite{Ronveaux} It has as a solution around the regular singular point $\xi=0$ given by the confluent Heun function\cite{Fiziev}
\begin{equation}
\label{heunsol}
	H_C(\alpha, \beta, \gamma, \delta, \eta, \xi) = \sum\limits_{n=0}^\infty v_n(\alpha, \beta, \gamma, \delta, \eta, \xi) \xi^n, \qquad
	\text{radius of convergence} \quad |\xi| < 1,
\end{equation}
where\cite{Mapping}
\begin{equation*}
\label{coeffs2}
 \delta = \mu + \nu - \frac{\alpha}{2}(\beta + \gamma + 2) = \frac{1}{4} U_0 d^2, \qquad \eta = \frac{\alpha}{2}(\beta + 1) - \mu - \frac{1}{2}(\beta + \gamma + \beta \gamma) = \frac{1}{4} (1 - (\varepsilon+U_0) d^2),
\end{equation*}
and the coefficients $v_n$ are given by the three-term recurrence relation\cite{Ronveaux}
\begin{equation}
\label{relation}
	A_n v_n = B_n v_{n-1} + C_n v_{n-2}, \qquad \text{with initial conditions} \qquad v_{-1} = 0, \qquad v_{0} = 1,
\end{equation}
where
\begin{subequations}
\label{recurrence}
 \begin{align}
  A_n &= 1 + \frac{\beta}{n}, \label{conda} \\
  B_n &= 1 + \frac{1}{n} \left( \beta + \gamma - \alpha - 1 \right) + 
				\frac{1}{n^2} \left\{ \eta - \frac{1}{2} (\beta + \gamma - \alpha) - \frac{\alpha \beta}{2} + \frac{\beta \gamma}{2} \right\}  , \label{condb} \\
  C_n &= \frac{\alpha}{n^2} \left( \frac{\delta}{\alpha} + \frac{\beta + \gamma}{2} + n - 1 \right). 
 \end{align}
\end{subequations}
Thus we have found the following solution to the Schr\"{o}dinger equation Eq.~\eqref{SE3}
\begin{equation}
\label{exactsol}
	\psi(\xi)_s = \xi^{\beta/2}~e^{\frac{\alpha}{2}\xi}~H_C( \alpha, \beta, \gamma, \delta, \eta, \xi).
\end{equation}
There is a general theorem of quantum mechanics that tells us when the Hamiltonian commutes with the parity operator, we should expect to find solutions of the Schr\"{o}dinger equation that are either symmetric or antisymmetric (at least for non-degenerate states). We can see from Eq.~\eqref{exactsol} that our solutions are wholly symmetric, as one would anticipate from the initial variable change to $\xi=1/\cosh^2(x/d)$. Therefore, we seek an antisymmetric solution using the more convenient odd variable $\zeta = \tanh(x/d)$,\cite{Hartmann} and we find
\begin{equation}
\label{exactsol2}
	\psi(\zeta)_a = \zeta (1-\zeta^2)^{\beta/2}~e^{-\frac{\alpha}{2}\zeta^2}~H_C( -\alpha, -\gamma, \beta, -\delta, \eta +\alpha^2/4, \zeta^2).
\end{equation}

In the next section, we shall look at the instances in which one can reduce the solution to a polynomial in $\xi$ or $\zeta$. Please note, a derivation similar to the one above can be carried for single well cases $p=-2, 0, 6$ and also in the second double-well case of $p=2$ for the $q=6$ family. The $q=4$ family is related by the fact in their respective confluent Heun equations the parameter $\alpha=0$ and parameters $\mu +\nu \ne 0$, frustrating termination attempts when employing the first termination condition. Whilst this problem does not arise for the $q=6$ family, the second termination condition cannot be satisfied, thus the confluent Heun functions arising in this case cannot be reduced to polynomials either, except for the special $(6, 4)$ case as we shall see.  

\section{\label{bound}Bound states in a hyperbolic double-well potential}

To reduce a confluent Heun function to a confluent Heun polynomial of degree $N$ we need two successive terms in the three-term recurrence relation Eq.~\eqref{relation} to vanish, halting the infinite series Eq.~\eqref{heunsol}. This requirement results in two termination conditions, which both need to be satisfied simultaneously\cite{Fiziev}
\begin{subequations}
\label{term}
 \begin{align}
  \mu + \nu + N \alpha = 0, \label{term1} \\
  \Delta_{N+1} (\mu) = 0. \label{term2}
 \end{align}
\end{subequations}
The first condition arises from ensuring $C_{N+2} = 0$, and is equivalent to $\frac{\delta}{\alpha} + \frac{\beta + \gamma}{2} + N + 1 = 0$ in the $(\delta, \eta)$ convention we use here, instead of the $(\mu, \nu)$ notation used by some other authors.\cite{Mapping} The second condition, which can be written as a tridiagonal determinant (please see Appendix~\ref{appendA}), arises from ensuring $v_{N+1} = 0$, such that it follows from Eq.~\eqref{relation} that all further terms in the series vanish identically. In our case, the first termination condition Eq.~\eqref{term1} allows us to find the following eigenvalue spectra
\begin{subequations}
\label{eigenvalues}
	\begin{align}
	\varepsilon_N^s = -\frac{1}{4 d^2} \left( 3 + 4N - d \sqrt{U_0} \right)^2, \qquad  U_0 d^2 >  (3 + 4N)^2, \\
	\varepsilon_N^a = -\frac{1}{4 d^2} \left( 5 + 4N - d \sqrt{U_0} \right)^2, \qquad  U_0 d^2 >  (3 + 5N)^2,
	\end{align}
\end{subequations}
for the symmetric (s) and antisymmetric (a) solutions respectively. The second termination condition Eq.~\eqref{term2} puts a constraint on the values the potential parameters $U_0$ and $d$ can take. We shall now give some illustrative examples of the first two states $N = 1, 2$.

\subsection{\label{sub1}The $N = 1$ state}

\begin{figure}[htbp]
 \begin{centering}
  \setlength{\unitlength}{1.0\textwidth}
  \includegraphics[width=0.65\textwidth]{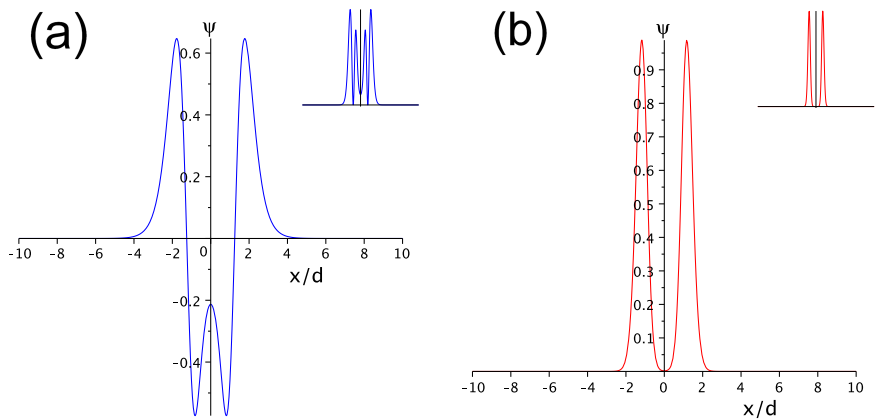}
 \end{centering}
 \caption{(Color online) Plots of two symmetric wavefunctions from the $N=1$ state, with $d=1$ and (a) $U_0 = 149.57...$ (blue line), (b) $U_0 = 595.84...$ (red line). Inset: associated probability density.}
 \label{fig:firstplot}
\end{figure}

\begin{figure}[htbp]
 \begin{centering}
  \setlength{\unitlength}{1.0\textwidth}
  \includegraphics[width=0.65\textwidth]{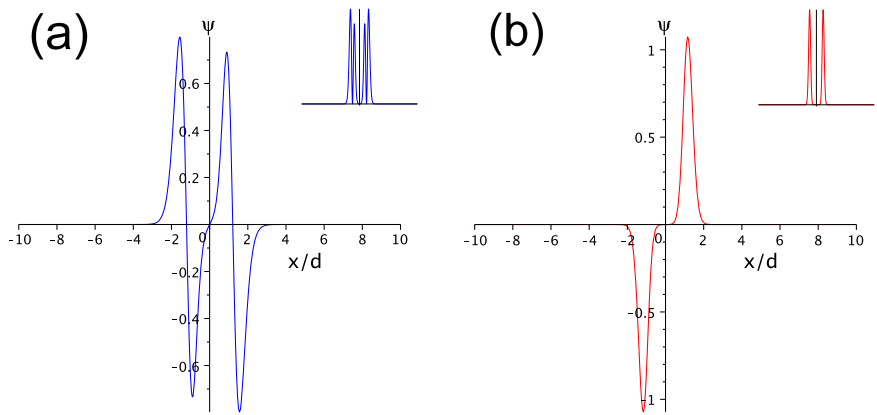}
 \end{centering}
 \caption{(Color online) Plots of two antisymmetric wavefunctions from the $N=1$ state, with $d=1$ and (a) $U_0 = 426.23...$ (blue line), (b) $U_0 = 1092.80...$ (red line). Inset: associated probability density.}
 \label{fig:oddfirstplot}
\end{figure}

Let us consider the first symmetric state, $N = 1$. Here the eigenvalue is $\varepsilon_1^s = -\frac{1}{4 d^2} \left( 7 - d \sqrt{U_0} \right)^2$, and we must ensure $U_0 d^2 > 49$ to guarantee Eq.~\eqref{term1} is satisfied. The second termination condition Eq.~\eqref{term2} can be re-written as the $2 \times 2$ matrix
\begin{equation}
\label{firstcond}
\begin{vmatrix}
 \mu - q_1 & 1+\beta \\
 \alpha & \mu - q_2 + \alpha
\end{vmatrix}
	= 0, \qquad \text{where} \qquad q_1 = 0, \qquad q_2 = 2 + \beta + \gamma
\end{equation}
which tells us what pair of values $U_0$ and $d$ can take. For simplicity, in what follows we choose $d = 1$, and upon solving Eq.~\eqref{firstcond} via root-finding methods we find the special potential strengths
\begin{equation}
\label{special1}
	U_0 = 149.57...,\quad \text{and} \quad 595.84...
\end{equation}
Thus, we have found the following wavefunction describes the first symmetric bound state in a hyperbolic double-well
\begin{equation}
\label{sol1}
	\psi(\xi)_s = e^{\frac{\alpha}{2}\xi}~\xi^{\beta/2}~\left( 1 + v_1 \xi \right).
\end{equation}

We plot Eq.~\eqref{sol1} and its associated probability density for the special values Eq.~\eqref{special1} in Fig.~\ref{fig:firstplot}, showing the symmetric parity of both wavefunctions as one would expect. We see that the higher the potential strength the tighter the confinement and a decrease in nodes from two to zero (although the factor $e^{\frac{\alpha}{2}\xi}$ in the wavefunction produces a superficial "node" at $x=0$ in the second case). Carrying out the same analysis for the first antisymmetric state, we find similar behavior but now with a decrease in nodes from three to one, as can be seen in Fig.~\ref{fig:oddfirstplot}.

\subsection{\label{sub2}The $N = 2$ state}

\begin{figure}[htbp]
 \begin{centering}
  \setlength{\unitlength}{1.0\textwidth}
  \includegraphics[width=0.95\textwidth]{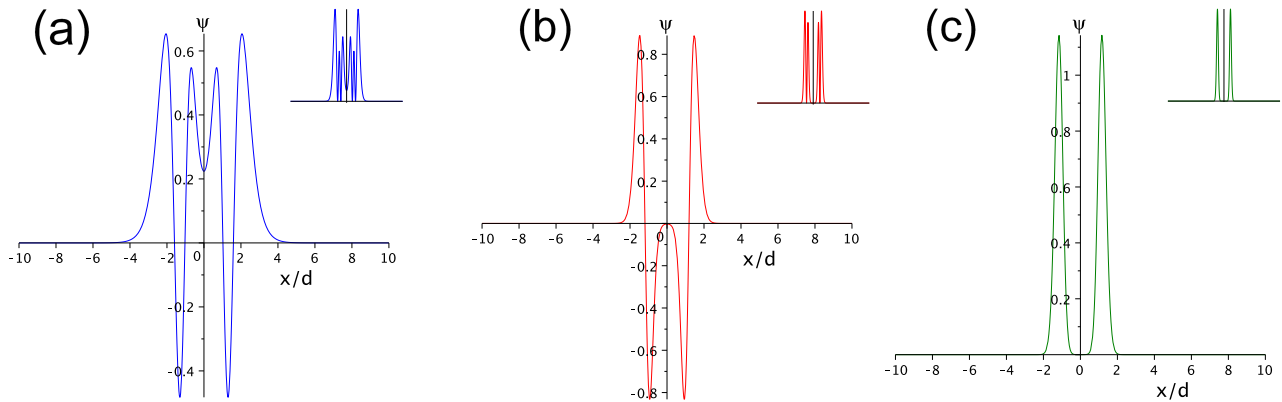}
 \end{centering}
 \caption{(Color online) Plots of three symmetric wavefunctions from the $N=2$ state, with $d=1$ and (a) $U_0 = 279.14...$ (blue line), (b) $U_0 = 860.32...$ (red line), (c) $U_0 = 2539.74...$ (green line). Inset: associated probability density.}
 \label{fig:secondplot}
\end{figure}

\begin{figure}[htbp]
 \begin{centering}
  \setlength{\unitlength}{1.0\textwidth}
  \includegraphics[width=0.95\textwidth]{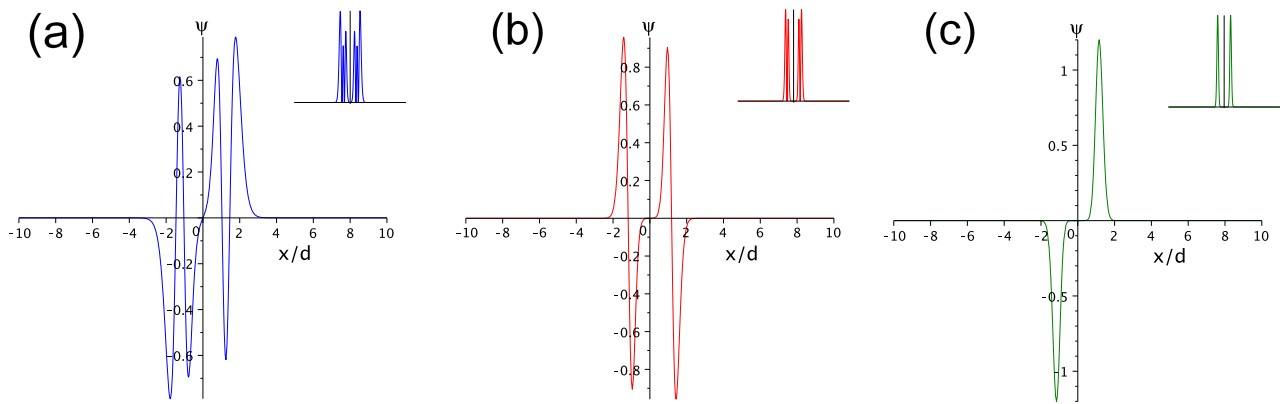}
 \end{centering}
 \caption{(Color online) Plots of three antisymmetric wavefunctions from the $N=2$ state, with $d=1$ and (a) $U_0 = 642.50...$ (blue line), (b) $U_0 = 1445.59...$ (red line), (c) $U_0 = 1740.79...$ (green line). Inset: associated probability density.}
 \label{fig:oddsecondplot}
\end{figure}

Proceeding as before, but this time for the $N = 2$ symmetric state, we immediately find from Eq.~\eqref{eigenvalues} the eigenvalue $\varepsilon_2^s = -\frac{1}{4 d^2} \left( 11 - d \sqrt{U_0} \right)^2$, subject to the stipulation $U_0 d^2 > 121$. The second termination condition Eq.~\eqref{term2} is equivalent to the $3 \times 3$ matrix
\begin{equation}
\label{secondcond}
\begin{vmatrix}
 \mu - q_1 & 1+\beta & 0 \\
 2\alpha & \mu - q_2 + \alpha & 2(2+\beta) \\
 0 & \alpha & \mu - q_3 + \alpha
\end{vmatrix}
	= 0, \qquad \text{where} \qquad q_1 = 0, \qquad q_2 = 2 + \beta + \gamma, \qquad q_3 = 2(3 + \beta + \gamma).
\end{equation}
Upon solving Eq.~\eqref{secondcond} we find the special values
\begin{equation}
\label{special2}
	U_0 = 279.14...,\quad 860.32...,\quad \text{and} \quad 1740.79...,
\end{equation}
again for $d = 1$. The form of the symmetric wavefunction follows straightforwardly from Eq.~\eqref{exactsol}
\begin{equation}
\label{sol2}
	\psi(\xi)_s = e^{\frac{\alpha}{2}\xi}~\xi^{\beta/2}~\left( 1 + v_1 \xi + v_2 \xi^2 \right).
\end{equation}
In Fig.~\ref{fig:secondplot}, we plot Eq.~\eqref{sol2} and its associated probability density for the special values Eq.~\eqref{special2}. It can be seen that the wavefunctions are symmetric, with a maximum of four nodes for the shallowest state, to two nodes for the middle state, to zero nodes for the deepest state. Again, the factor $e^{\frac{\alpha}{2}\xi}$ in the wavefunction creates superficial "nodes" at $x=0$ in the second and third cases. The equivalent antisymmetric solution is displayed in Fig.~\ref{fig:oddsecondplot}, illustrating the anticipated pattern: a drop from five nodes to three nodes to one node as we increase the potential strength.

Higher states $N=2, 3, ...$ can be obtained by following the same recipe. For wavefunctions of a state $N$, there are $N+1$ solutions with a different $U_0$ for a certain $d$. The symmetric wavefunction confined in the lowest potential strength $U_0$ will have $2 N$ nodes, and the node number will decrease by two each time until the deepest symmetric state has no nodes. The antisymmetric wavefunction confined in the lowest potential strength will have the most nodes, $2 N + 1$, which decreases by two each time until the deepest state has just one node. As the potential is symmetric, we find as expected the parity of the polynomial solutions changes alternately from even to odd as we increase the potential strength and hit successively higher eigenvalues.

\section{\label{disc}Discussion}

\begin{figure}[htbp]
 \begin{centering}
  \setlength{\unitlength}{1.0\textwidth}
  \includegraphics[width=0.95\textwidth]{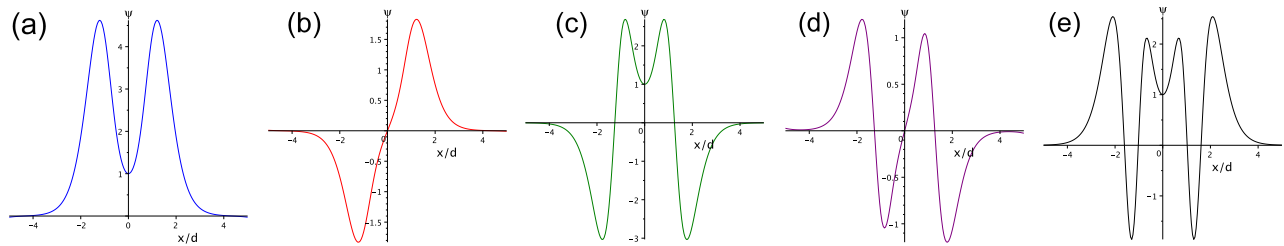}
 \end{centering}
 \caption{(Color online) Plots of the evolution of the unnormalized bound state wavefunctions from ground state to fourth excited state, with $d=1$ and (a) $U_0 = 71.46...$ (blue line), (b) $U_0 = 72.51...$ (red line), (c) $U_0 =  149.57...$ (green line), (d) $U_0 = 152.50...$ (purple line) and (d) $U_0 =  258.38...$ (black line). We chose $\varepsilon = -6.84$ such that plot (c) corresponds to a close-form result, namely that of Fig.~\ref{fig:firstplot} (a). }
 \label{fig:numeric}
\end{figure}

We have seen the existence of bound states associated to confluent Heun polynomial wavefunctions at special values of the system parameters. The Schr\"{o}dinger equation admits other bound states, such as a ground state, which can be generated by numeric integration. There seems to be no qualitative difference between bound states at 'special' values, coupled to closed-form wavefunctions, and other values, found from numeric integration. We show in Fig.~\ref{fig:numeric} how modulation of $U_0$ brings one from the ground state to excited states, which in this case includes a second excited state which corresponds to a close-form result, namely that of Fig.~\ref{fig:firstplot} (a).

\section{\label{conc}Conclusion}

We have reported a family of confining hyperbolic potentials which allow one to transform the one-dimensional Schr\"{o}dinger equation to a confluent Heun equation. One case, describing a hyperbolic double-well, can be further reduced such that the wavefunctions associated with bound states can be written in terms of confluent Heun polynomials. We expect this work to be of general interest due its simplicity and focus on the well-known double-well problem of quantum mechanics, but also to be intriguing to those interested in the use of the exotic but increasingly popular Heun differential equation in physics.

\section*{Acknowledgments}
We would like to thank M.~E.~Portnoi and N.~Tufnel for useful discussions related to the nature of the solutions and A.~M.~Alexeev and L.~Marnham for a critical reading of the manuscript. This work was supported by the EPSRC.

\begin{appendix}

\section{\label{appendA}The second termination condition as a tridiagonal determinant}

The second termination condition, $\Delta_{N+1} (\mu) = 0$, can be represented as the following tridiagonal determinant\cite{Fiziev}
\begin{equation}
\label{large}
\begin{vmatrix}
 \mu - q_1 & (1+\beta) & 0 & \dots & 0 & 0 & 0 \\
  N \alpha & \mu - q_2 + \alpha & 2(2+\beta) & \dots & 0 & 0 & 0\\
   0 & (N-1)\alpha & \mu-q_3+ 2\alpha & \dots & 0 & 0 & 0\\
    \vdots & \vdots & \vdots & \ddots & \vdots & \vdots & \vdots\\
     0 & 0 & 0 & \dots & \mu - q_{N-1} + (N-2)\alpha & (N-1)(N-1+\beta) & 0\\
      0 & 0 & 0 & \dots & 2\alpha & \mu - q_N + (N-1)\alpha & N(N+\beta)\\
      0 & 0 & 0 & \dots & 0 & \alpha & \mu - q_{N+1} + N \alpha \\
\end{vmatrix}
	= 0, 
\end{equation}
where $q_n = (n-1)(n+ \beta + \gamma)$.

\end{appendix}


\begin{thebibliography}{100}
\bibitem{Mapping}The reverse mapping is given by $\mu = \frac{1}{2}(\alpha - \beta -\gamma + \alpha \beta - \beta \gamma) - \eta$ and $\nu = \frac{1}{2}(\alpha + \beta + \gamma + \alpha \gamma + \beta \gamma) + \delta + \eta$. Please note, there are two common parameterizations of the confluent Heun equation: the $(\delta, \eta)$ parameters are used in Ref.~[\onlinecite{Ronveaux}], whereas the alternative $(\mu, \nu)$ parameters were introduced in Ref.~[\onlinecite{Fiziev}].
\end{thebibliography}
\end{document}